\documentclass[manuscript]{aastex62}

\usepackage{txfonts}
\usepackage{latexsym,bm}
\usepackage{url}
\usepackage{color}
\usepackage{colortbl}
\usepackage{multirow}

\definecolor{mygray}{gray}{.9}

\shorttitle{DIFFUSION WITH DISPLACEMENT VARIANCE}
\shortauthors{WANG AND QIN}

\begin{document}
              \arraycolsep 0pt

\title{DIFFUSION COEFFICIENT WITH DISPLACEMENT VARIANCE
OF ENERGETIC PARTICLES WITH ADIABATIC FOCUSING}

\correspondingauthor{G. Qin}
\email{qingang@hit.edu.cn}

\author[0000-0002-9586-093X]{J. F. Wang}
\affiliation{School of Science, Harbin Institute of Technology, Shenzhen,
518055, China; qingang@hit.edu.cn}

\author[0000-0002-3437-3716]{G. Qin}
\affiliation{School of Science, Harbin Institute of Technology, Shenzhen,
518055, China; qingang@hit.edu.cn}

\begin{abstract}
The equation $\kappa_{zz}=d\sigma^2/(2dt)$
(hereafter DCDV)
is a well-known formula
of energetic particles
describing the relation
of parallel diffusion
coefficient $\kappa_{zz}$
with the parallel displacement
variance $\sigma^2$.
In this study, we find that
DCDV
is only applicable to
two kinds of transport equations
of isotropic distribution function,
one is
without cross terms,
the other is
without convection term.
Here,
by employing the more general transport
equation,
i.e., the variable coefficient
differential equation
derived from the Fokker-Planck
equation,
a new equation of $\kappa_{zz}$
as a function of $\sigma^2$ is obtained.
We find that
DCDV
is the special case of the new equation.
In addition, another equation of
$\kappa_{zz}$ as a function of $\sigma^2$
corresponding to
the telegraph equation is also
investigated preliminarily.

\end{abstract}


\section{INTRODUCTION}

It is very interesting to study
energetic charged particle
diffusion in
the magnetic turbulence superposed on
the mean magnetic field
in astrophysics
 (e.g.,
cosmic ray physics, astrophysical
plasmas, and
space weather research)
and fusion plasma physics
\citep{Jokipii1966,
Schlickeiser2002, MatthaeusEA2003,
ShalchiEA2005, ShalchiEA2006, Qin2007,
HauffEA2008, Shalchi2009a,
Shalchi2010, QinEA2014}.
The interaction process between particles
and turbulent magnetic fields is very
complicated, and
particles and magnetic field lines
have the stochastic
properties. Therefore, one has to employ
methods of statistics to describe
the complicated transport of energetic particles
\citep{Earl1974, Earl1976,
BeeckEA1986, Shalchi2011,
Litvinenko2012a,
Litvinenko2012b, ShalchiEA2013, HeEA2014,
WangEA2017a, WangEA2017b, wq2018}
and field line random walk
\citep{MatthaeusEA1995,
ShalchiAKourakis2007,
ShalchiAQin2010}.
Because the background
magnetic field $\vec{B}_0$
breaks
the symmetry of the
magnetized plasma,
one has to distinguish
particle diffusion
along and across
the large-scale magnetic field.
One may
only consider the parallel
diffusion in many cases, when it is
much greater than the perpendicular
one
\citep[see, e.g.,][]{Earl1974, Earl1976,
BeeckEA1986, Shalchi2009b, Shalchi2011,
Litvinenko2012a,
Litvinenko2012b, ShalchiEA2013,
HeEA2014, wq2018}.

Various analytical theories of
parallel diffusion for energetic
charged particles have been
developed in the past.
The first attempt to resolve the parallel
diffusion was
the development of quasilinear theory
\citep{Jokipii1966} which corresponds
to the first order
perturbation theory.
Quasilinear theory, however,
is problematic and usually does
not agree with test-particle simulations
\citep{QinEA2002a, QinEA2002b,
Shalchi2009a}.
With the second order quasilinear theory
(SOQLT) developed by \citet{Shalchi2005},
by using the Earl's formula
\citep{Earl1974}
the parallel
diffusion coefficient
can be evaluated
more accurately
\citep{ShalchiEA2009a,
ShalchiEA2009b, ReimerEA2016}.
On the other hand,
by employing the assumptions
and approximations
used in NonLinear Guiding Center (NLGC)
theory \citep{MatthaeusEA2003},
\citet{Qin2007} developed an extended
parallel diffusion theory.

The mean square displacement
$\sigma^2=\langle (\Delta z)^2 \rangle
-\langle (\Delta z) \rangle^2$
is defined with the first and second order parallel
displacement moments of the isotropic
distribution function
of the energetic particles.
$\sigma^2$, describing the spread of
possible particle orbits,
is one of the fundamental quantities in the parallel transport.
In previous research
\citep[see, e.g.,][]{Shalchi2009a}
the parallel diffusion
coefficient $\kappa_{zz}$
is usually described by
the following formula
(hereafter DCDV)
\begin{equation}
\kappa_{zz}=\frac{1}{2}\frac{d\sigma^2}{dt}.
\label{kzz-sigma}
\end{equation}

To study energetic particle diffusion,
the background magnetic field $\vec{B}_0$,
e.g., mean solar wind magnetic field
in the interplanetary space,
is usually considered as a constant.
However,
it is clear that the mean solar wind
magnetic field
is not constant in reality,
especially when particles are
close to the Sun.
It is found
that the
spatially varying
background solar wind magnetic fields lead
to the adiabatic focusing effect of
charged energetic particle
transport and introduce
correction to the particle
diffusion coefficients
\citep[see, e.g.,][]{Roelof1969,
Earl1976, Kunstmann1979,BeeckEA1986,
BieberEA1990,
Kota2000, SchlickeiserEA2008,
Shalchi2009b, Shalchi2011,
Litvinenko2012a,
Litvinenko2012b, ShalchiEA2013,
WangEA2016, WangEA2017b, wq2018}.
The adiabatic focusing effect
causes
a convection term in
the energetic particles transport equation
of the isotropic distribution function,
so DCDV might be modified.

In order to address this question,
in this paper we derive a new formula
for the parallel diffusion coefficient
with the parallel displacement variance
by modifying DCDV with the
adiabatic focusing
effect.
The paper is organized as follows.
In Section \ref{EQUATION OF ISOTROPIC
DISTRIBUTION FUNCTION},
from the Fokker-Planck equation
with adiabatic focusing effect
 we derive
the isotropic
distribution function equation
and the telegraph equation.
In Section \ref{THE PARALLEL DIFFUSION
COEFFICIENT REPRESENTED BY
ALONG-FIELD DISPLACEMENT MOMENTS},
we evaluate the first
and second order moments of
the parallel displacement
to derive the modified DCDV.
In Section \ref{SPECIAL CASES},
some special cases of
the modified  DCDV
are explored.
We conclude and summarize
our results in Section
\ref{SUMMARY AND CONCLUSION}.

\section{EQUATION OF ISOTROPIC
DISTRIBUTION FUNCTION}
\label{EQUATION OF ISOTROPIC
DISTRIBUTION FUNCTION}

The standard Fokker-Planck equation,
which incorporates the effects
of pitch-angle
scattering and along-field
adiabatic focusing,
is given as
\citep{Roelof1969, Earl1981}
\begin{equation}
\frac{\partial{f_0}}{\partial{t}}
+ v\mu
\frac{\partial{f_0}}
{\partial{z}}
=\frac{\partial{}}
{\partial{\mu}}
\left[D_{\mu \mu}(\mu)
\frac{\partial{f_0}}
{\partial{\mu}}\right]
-\frac{v}{2L}(1-\mu^2)
\frac{\partial{f_0 }}
{\partial{\mu}},
\label{standard Fokker-Planck
equation}
\end{equation}
which is usually used
in previous research
\citep{SchlickeiserEA2007,
Shalchi2011,
LitvinenkoASchlickeiser2013,
EffenbergerEA2014,
MalkovSagdeev2015,
WangEA2016}.
Here $f_0$ is the isotropic
distribution function,
$t$ is time, $z$
is the distance
along the background
magnetic field,
$\mu=v_z /v$ is the pitch-angle
cosine with particle
speed $v$ and
its z-component $v_z$,
$D_{\mu \mu}(\mu)$
is the
pitch-angle diffusion
coefficient,
$L(z)=-B_0 (z)/ [dB_0 (z) / dz]$
is the adiabatic focusing
characteristic length of the
large-scale magnetic field $B_0(z)$.
In this paper,
for simplification
we assume that
the  adiabatic focusing
characteristic length
is a constant.
The source term and
the terms related to
momentum diffusion
and so on are ignored
in Equation
(\ref{standard Fokker-Planck
equation}).
The more complete form of
the Fokker-Planck equation
can be found in
\citet{Schlickeiser2002}.

By introducing the linear
phase space density
$f(z,\mu,t)=f_0(z,\mu,t)/B_0$,
from Equation
(\ref{standard Fokker-Planck
equation}),
the modified Fokker-Planck equation
for the
distribution function of energetic
charged particles can be obtained
\citep{Kunstmann1979,HeEA2014,
WangEA2017b, wq2018}
\begin{equation}
\frac{\partial{f}}{\partial{t}}
+ v\mu
\frac{\partial{f}}
{\partial{z}}=
\frac{\partial{}}{\partial{\mu}}
\left[D_{\mu \mu}(\mu)
\frac{\partial{f}}
{\partial{\mu}}-\frac{v}{2L}
(1-\mu^2)f \right].
\label{Fokker-Planck equation-0}
\end{equation}
Here, $D_{\mu\mu}(\mu)$ is time-independent. However,
if the temporal characteristic
of the pitch angle diffusion
need to be considered, $D_{\mu\mu}=D_{\mu\mu}(\mu,t)$,
Equation
(\ref{Fokker-Planck equation-0})
can be generalized as
\begin{equation}
\frac{\partial{f}}{\partial{t}}
+ v\mu
\frac{\partial{f}}
{\partial{z}}=
\frac{\partial{}}{\partial{\mu}}
\left[D_{\mu \mu}(\mu,t)
\frac{\partial{f}}
{\partial{\mu}}-\frac{v}{2L}
(1-\mu^2)f \right].
\label{Fokker-Planck equation}
\end{equation}
In order for the pitch angle diffusion equation to hold,
the characteristic
time of the pitch angle
diffusion has to
be much less than
the characteristic
time of $D_{\mu\mu}(\mu,t)$.
Equation (\ref{Fokker-Planck equation}) is the starting
point of the research in
this paper.

With strong pitch-angle scattering,
the gyro-tropic
cosmic-ray phase space
density $f(\vec{x}, \mu, t)$
can be split into
the dominant isotropic part
$F(\vec{x}, t)$
and the subordinate
anisotropic part
$g(\vec{x}, \mu, t)$
\citep[see, e.g.,][]
{SchlickeiserEA2007,
SchlickeiserEA2008, HeEA2014,
WangEA2017b, wq2018}
\begin{equation}
f(\vec{x}, \mu, t)=F(\vec{x}, t)
+g(\vec{x}, \mu, t)
\end{equation}
with
\begin{equation}
F(\vec{x}, t)=\frac{1}{2}
\int_{-1}^1 d\mu
f(\vec{x}, \mu, t)
\end{equation}
and
\begin{equation}
\int_{-1}^1 d\mu
g(\vec{x}, \mu, t)=0 .
\end{equation}

\subsection{The variable
coefficient differential
equation of the
isotropic distribution function
$F(z, t)$}
\label{The governing equation
of the
isotropic distribution function
F}

In this subsection,
by employing the method
in \citet{wq2018}
we derive
the formula of the variable
coefficient differential
equation of the
isotropic distribution function
$F(z, t)$, and the derivation are
similar to Equation (22) in
\citet{wq2018}.

Integration of Equation
 (\ref{Fokker-Planck equation})
over $\mu$ gives
\begin{equation}
\frac{\partial{F}}{\partial{t}}
+ \frac{v}{2}
\frac{\partial{}}
{\partial{z}}\int_{-1}^{1}
\mu g d\mu=0.
\label{Equation of F with g}
\end{equation}
Next, integrating Equation
(\ref{Fokker-Planck equation})
over $\mu$ from $-1$ to $\mu$,
and using the regularity
$D_{\mu\mu}(\mu=\pm 1)=0$
we can find
\begin{eqnarray}
\frac{\partial{F}}
{\partial{t}}(\mu+1)
&&+ \frac{\partial{}}
{\partial{t}}
\int_{-1}^{\mu}d\nu g
+\frac{v(\mu^2-1)}{2}
\frac{\partial{F}}
{\partial{z}}
+v\frac{\partial{}}
{\partial{z}}
\int_{-1}^{\mu}
d\nu \nu g 
=D_{\mu\mu}(\mu,t)
\frac{\partial{g}}
{\partial{\mu}}
-\frac{v(1-\mu^2)}{2L}F
-\frac{v(1-\mu^2)}{2L}g.
\label{integrate
from -1 to mu}
\end{eqnarray}
By subtracting Equation
(\ref{Equation of F with g})
from
(\ref{integrate from -1 to mu})
we can find the
following equation
\begin{eqnarray}
\frac{\partial{F}}
{\partial{t}}\mu
+ \frac{\partial{}}
{\partial{t}}
\int_{-1}^{\mu}d\nu g
&&+\frac{v(\mu^2-1)}{2}
\frac{\partial{F}}
{\partial{z}}
+v\frac{\partial{}}
{\partial{z}}
\int_{-1}^{\mu}
d\nu \nu g 
-\frac{v}{2}
\frac{\partial{}}
{\partial{z}}\int_{-1}^{1}
\mu g d\mu \nonumber\\
&&=D_{\mu\mu}(\mu,t)
\frac{\partial{g}}
{\partial{\mu}}
-\frac{v(1-\mu^2)}{2L}F
-\frac{v(1-\mu^2)}{2L}g.
\label{integrate from
-1 to mu -2}
\end{eqnarray}
After a straightforward
algebra, Equation
(\ref{integrate from -1 to mu -2})
reduces to the following form
\begin{equation}
\frac{\partial{g}}
{\partial{\mu}}-
\frac{v(1-\mu^2)g}
{2L D_{\mu \mu}(\mu,t)}
+\frac{v(1-
\mu^2)}{2D_{\mu \mu}(\mu,t)}
\left(\frac{\partial{F}}
{\partial{z}}
-\frac{F}
{L}\right)=\Phi(\mu,t)
\label{Equation of g with phi}
\end{equation}
with
\begin{eqnarray}
\Phi(\mu,t)=
\frac{1}{D_{\mu\mu}(\mu,t)}
\Bigg[\left(\frac{\partial{F}}
{\partial{t}}\mu
+\frac{\partial{}}{\partial{t}}
\int_{-1}^{\mu}gd\nu\right)
+\frac{v}{2}\frac{\partial{}}
{\partial{z}}
\left(2\int_{-1}^{\mu}d\nu \nu g-
\int_{-1}^{1}d\mu \mu g\right)\Bigg].
\label{Phi}
\end{eqnarray}
Equation
(\ref{Equation of g with phi})
can be rewritten as
\begin{equation}
\frac{\partial{}}{\partial{\mu}}
\Bigg\{\Bigg[g(\mu,t)
-L\left(\frac{\partial{F}}
{\partial{z}}
-\frac{F}{L} \right)
\Bigg]e^{-M(\mu,t)}\Bigg\}
=e^{-M(\mu,t)}\Phi(\mu,t)
\label{HS-like}
\end{equation}
with
\begin{equation}
M(\mu,t)=\frac{v}{2L}
\int_{-1}^{\mu} d\nu
\frac{1-\nu^2}{D_{\nu \nu}(\nu,t)}.
\label{M(mu)}
\end{equation}

In additon, with the same procedure used
in \citet{wq2018},
the anisotropic
distribution function can be
obtained as following
\begin{equation}
g(\mu,t)=L\left(\frac{\partial{F}}
{\partial{z}}
-\frac{F}{L}\right)\left[1-
\frac{2e^{M(\mu,t)}}{\int_{-1}^{1}
d\mu e^{M(\mu,t) }}\right]
+e^{M(\mu,t)}\left[R(\mu,t)
-\frac{\int_{-1}^{1}d\mu
e^{M(\mu,t)}R(\mu,t)}
{\int_{-1}^{1}d\mu
e^{M(\mu,t) }}\right]
\label{g}
\end{equation}
with
\begin{eqnarray}
R(\mu,t)&=&\int_{-1}^{\mu} d\nu
e^{-M(\nu,t)}\Phi(\nu,t).
\label{R(mu)}
\end{eqnarray}
Equations (\ref{g}) and (\ref{R(mu)})
contain
the effect from the term on the
right hand side of
Equation (\ref{HS-like}).
Combining Equations (\ref{Phi}),
(\ref{g}) and  (\ref{R(mu)})
gives the iterative function of
$g(\mu,t)$,
\begin{eqnarray}
g(\mu,t)&=&L\left(\frac{\partial{F}}
{\partial{z}}
-\frac{F}{L}\right)\left[1-
\frac{2e^{M(\mu,t)}}{\int_{-1}^{1}
d\mu e^{M(\mu,t) }}\right]
+e^{M(\mu,t)}\Bigg\{\int_{-1}^{\mu}
d\nu
e^{-M(\nu,t)}
\frac{1}{D_{\nu\nu}(\nu,t)}
\Bigg[\left(\frac{\partial{F}}
{\partial{t}}\nu
+\frac{\partial{}}{\partial{t}}
\int_{-1}^{\nu}
g(\rho,t)d\rho\right)
\nonumber \\
&&+\frac{v}{2}\frac{\partial{}}
{\partial{z}}
\left(2\int_{-1}^{\nu}d\rho \rho
g(\rho,t)-
\int_{-1}^{1}d\mu \mu
g(\mu,t)\right)
\Bigg]\nonumber\\
&&-\frac{1}
{\int_{-1}^{1}d\mu
e^{M(\mu,t) }}\int_{-1}^{1}d\mu
e^{M(\mu,t)}\int_{-1}^{\mu} d\nu
e^{-M(\nu,t)}
\frac{1}{D_{\nu\nu}(\nu,t)}
\Bigg[\left(\frac{\partial{F}}
{\partial{t}}\nu
+\frac{\partial{}}{\partial{t}}
\int_{-1}^{\nu}g(\rho,t)
d\rho\right)\nonumber\\
&&+\frac{v}{2}
\frac{\partial{}}
{\partial{z}}
\left(2\int_{-1}^{\mu}
d\nu \nu g(\nu,t)-
\int_{-1}^{1}d\mu \mu
g(\mu,t)\right)\Bigg]\Bigg\}.
\label{g2}
\end{eqnarray}
By iterating Equation
(\ref{g2}),
we obtain the formula of
$g(\mu,t)$ as
\begin{equation}
g(\mu,t)=\sum_{m, n}
\epsilon_{m,n}(t)
\frac{\partial^{m+n}{}}
{\partial {t^m}
\partial{z}^n}  F
\label{expanded g}
\end{equation}
with the time dependent coefficients
$\epsilon_{m,n}(t)$.
Here $m,n=0,1,2,3,\cdots$.
Furthermore,
considering Equation
(\ref{Phi}), (\ref{g})
and Equation (\ref{R(mu)}),
we can write
$R(\mu,t)$ as
\begin{equation}
R(\mu,t)=\sum_{m, n}
\chi_{m,n}(t)
\frac{\partial^{m+n}{}}
{\partial {t^m}
\partial{z}^n}  F
\label{series of R}
\end{equation}
with the time dependent coefficients
$\chi_{m,n}(t)$ and
$m,n=0,1,2,3,\cdots$.
The coefficients
$\chi_{m,n}(t)$
in the latter equation
is related to
the coefficients
$\epsilon_{m,n}(t)$
in Equation
(\ref{expanded g}).

In order to
obtain the differential equation
of isotropic distribution
function,
we need to get the
following formula
by using Equation
(\ref{Equation of F with g})
\begin{equation}
\int_{-1}^{1}d\mu \mu g(\mu,t)
=-2\frac{\int_{-1}^{1}d\mu\mu
e^{M(\mu,t)}}
{\int_{-1}^{1}d\mu e^{M(\mu,t)}}
\left(\frac{\partial{F}}
{\partial{z}}
-\frac{F}{L}\right)L
+\int_{-1}^{1}
d\mu \mu e^{M(\mu,t)}
\Bigg[R(\mu,t)
-\frac{\int_{-1}^{1}d\mu
e^{M(\mu,t)}R(\mu,t)}
{\int_{-1}^{1}d\mu
e^{M(\mu,t)}}
\Bigg].
\label{Integrating mu g over
mu from -1 to 1}
\end{equation}
By inserting Equation
(\ref{Integrating mu g over
mu from -1 to 1})
into Equation
(\ref{Equation of F with g})
one can obtain
\begin{eqnarray}
\frac{\partial{F}}{\partial{t}}
&-&\frac{\partial{}}
{\partial{z}}
\left[vL\frac{\int_{-1}^{1}
d\mu \mu e^{M(\mu,t)}}
{\int_{-1}^{1}d\mu
e^{M(\mu,t)}}
\left(\frac{\partial{F}}
{\partial{z}}-
\frac{F}{L}\right)\right]
+\Lambda(z,t)=0
\label{Accurate
diffusion equation}
\end{eqnarray}
with
\begin{equation}
\Lambda(z,t)=
\frac{v}{2}\int_{-1}^{1}
d\mu \mu e^{M(\mu,t)}
\left[\frac{\partial{R(\mu,t)}}
{\partial{z}}
-\frac{\int_{-1}^{1}d\mu
\frac{\partial{R(\mu,t)}}
{\partial{z}}
e^{M(\mu,t)}}
{\int_{-1}^{1}d\mu
e^{M(\mu,t)}}
\right].
\label{Lambda}
\end{equation}
By inserting Equation
(\ref{series of R})
into Equation (\ref{Lambda})
we can find
\begin{equation}
\Lambda(z,t)
=\sum_{m,n}\eta_{m,n}(t)
\frac{\partial^{m+n}}
{\partial {t^m}
\partial{z}^n}  F
\label{Lambda-eta}
\end{equation}
with the coefficients
$\eta_{m,n}(t)$ and
$m,n=0,1,2,3,\cdots$.
Thus, Equation
(\ref{Accurate diffusion
equation}) can be written as
\begin{eqnarray}
\frac{\partial{F}}
{\partial{t}}
-\frac{\partial{}}
{\partial{z}}&&
\left[vL\frac{\int_{-1}^{1}
d\mu \mu e^{M(\mu, t)}}
{\int_{-1}^{1}d\mu
e^{M(\mu,t)}}
\left(\frac{\partial{F}}
{\partial{z}}-
\frac{F}{L}\right)\right]
+\sum_{m,n}\eta_{m,n}(t)
\frac{\partial^{m+n}}
{\partial {t^m}
\partial{z}^n} F=0.
\label{Accurate diffusion
equation-2}
\end{eqnarray}
Now we get
a variable coefficient
linear differential equation
of the isotropic distribution
function.

\subsection{The specific form of the variable coefficient differential equation}

By employing the method in
\citet{wq2018},
one can rewrite Equation
(\ref{Accurate diffusion
	equation-2})
as
\begin{eqnarray}
\frac{\partial{F}}
{\partial{t}}=
&&-\kappa_z(t)
\frac{\partial{F}}{\partial{z}}
+\kappa_{zz}(t)
\frac{\partial^2{F}}{\partial{z^2}}
+\left(\kappa_{zzz}(t)
\frac{\partial^3{F}}{\partial{z^3}}
+\kappa_{zzzz}(t)
\frac{\partial^4{F}}{\partial{z^4}}
+\cdots\right)
+\left( \kappa_{tz}(t)
\frac{\partial^2{F}}{\partial{t}
\partial{z}}
+ \kappa_{ttz}(t)
\frac{\partial^3{F}}
{\partial{t^2}\partial{z}}
+ \kappa_{tttz}(t)
\frac{\partial^4{F}}{\partial{t^3}
\partial{z}}+\cdots\right)
\nonumber\\
&&+\left( \kappa_{tzz}(t)
\frac{\partial^3{F}}{\partial{t}
\partial{z^2}}
+ \kappa_{ttzz}(t)\frac{\partial^4{F}}
{\partial{t^2}\partial{z^2}}
+ \kappa_{tttzz}(t)\frac{\partial^5{F}}
{\partial{t^3}\partial{z^2}}
+\cdots\right)
+\cdots\cdots,
\label{equation of F}
\end{eqnarray}
here, $\kappa_z(t)$,
$\kappa_{zz}(t)$,
$\kappa_{zzz}(t)$,
$\kappa_{zzzz}(t)$,
$\cdots\cdots$,
$\kappa_{tz}(t)$,
$\kappa_{ttz}(t)$, $\kappa_{tttz}(t)$,
$\cdots\cdots$,
$\kappa_{tzz}(t)$, $\kappa_{ttzz}(t)$,
$\kappa_{tttzz}(t)$,
$\cdots\cdots$
are all time dependent.
This equation determines the transport regimes by the specific
forms of its coefficient.
For example, as shown in
\citet{wq2018},
$\kappa_z(t)$
is the coefficient of the convection
term
\begin{equation}
\kappa_z(t)
=v\frac{\int_{-1}^{1}d\mu
\mu e^{M(\mu,t)}}{\int_{-1}^{1}
d\mu e^{M(\mu,t)}},
\label{kz}
\end{equation}
and $\kappa_{zz}(t)$
is the coefficient of the parallel
diffusion term
\begin{eqnarray}
\kappa_{zz}(t)&=&
vL\frac{\int_{-1}^{1}d\mu
\mu e^{M(\mu,t)}}{\int_{-1}^{1}
d\mu e^{M(\mu,t)}}
+\frac{v^2}{2}
\Bigg\{
\int_{-1}^{1}d\mu
\mu e^{M(\mu,t)}
\int_{-1}^{\mu}d\nu
\frac{e^{-M(\nu,t)}}
{D_{\nu \nu}(\nu,t)}
\left[\int_{-1}^{\nu}d\rho\rho
\left(1-\frac{2e^{M(\rho,t)}}
{\int_{-1}^{1}d\mu
e^{M(\mu,t)}}\right)
+\int_{-1}^{1} d\mu \mu
\frac{e^{M(\mu,t)}}
{\int_{-1}^{1} d\mu
e^{M(\mu,t)}}\right]
\nonumber\\
&&-\frac{\int_{-1}^{1}
d\mu \mu
e^{M(\mu,t)}}{\int_{-1}^{1}
d\mu e^{M(\mu,t)}}
\int_{-1}^{1}
d\mu e^{M(\mu,t)}
\int_{-1}^{\mu}
d\nu \frac{e^{-M(\nu,t)}}
{D_{\nu \nu}(\nu,t)}
\left[\int_{-1}^{\nu}
d\rho\rho
\left(1-\frac{2e^{M(\rho,t)}}
{\int_{-1}^{1}d\mu
e^{M(\mu,t)}}\right)
+\int_{-1}^{1} d\mu \mu
\frac{e^{M(\mu,t)}}
{\int_{-1}^{1} d\mu
e^{M(\mu,t)}}\right]
\Bigg\}.
 \label{A1-iso}
\end{eqnarray}
We can similarly obtain other
time-dependent coefficients in Equation
(\ref{equation of F}).

\subsection{Telegraph equation of the
isotropic distribution function
$F(z, t)$}
\label{Telegraph equation of the
isotropic distribution function}

The diffusion convection equation
is frequently used in previous research
\citep[see, e.g.,][]{Shalchi2009a}.
But the accuracy of the diffusion
approximation is limited
because the signal speed is
infinite in the diffusion limit.
Recently, an improved description
of the
energetic particle transport
is provided by the telegraph equation,
characterized by a finite
signal propagation speed
\citep{Kota1994,
PaulsBurger1994,
SchwadronEA2010, PorraMasoliver1997,
ZankEA2000, FedorovShakhov2003,
KaghashviliEA2004,
LitvinenkoASchlickeiser2013,
EffenbergerEA2014, LitvinenkoEA2015,
LitvinenkoNoble2016}.
Using the iterative method
in \citet{LitvinenkoASchlickeiser2013},
one can obtain the
telegraph equation from
Equation (\ref{equation of F})
\begin{eqnarray}
\frac{\partial{F}}{\partial{t}}
+\tau(t)\frac{\partial^2{F}}{\partial{t^2}
}+\kappa_z(t)\frac{\partial{F}}
{\partial{z}}=\kappa_{zz}(t)
\frac{\partial^2{F}}{\partial{z^2}}
\label{telegraph equation}
\end{eqnarray}
with
\begin{eqnarray}
\tau(t)=-\frac{\kappa_{tzz}(t)}
{\kappa_{zz}(t)}.
\label{tau}
\end{eqnarray}

Equation (\ref{equation of F})
can be rewritten as
\begin{eqnarray}
\frac{\partial{F}}{\partial{z}}
=&&\frac{1}{\kappa_z(t)}
\Bigg[-\frac{\partial{F}}
{\partial{t}}
+\kappa_{zz}(t)
\frac{\partial^2{F}}{\partial{z^2}}
+\left(\kappa_{zzz}(t)
\frac{\partial^3{F}}{\partial{z^3}}
+\kappa_{zzzz}(t)
\frac{\partial^4{F}}{\partial{z^4}}
+\cdots\right)
+\left( \kappa_{tz}(t)
\frac{\partial^2{F}}{\partial{t}
	\partial{z}}
+ \kappa_{ttz}(t)
\frac{\partial^3{F}}
{\partial{t^2}\partial{z}}
+ \kappa_{tttz}(t)
\frac{\partial^4{F}}{\partial{t^3}
	\partial{z}}+\cdots\right)
\nonumber\\
&&+\left( \kappa_{tzz}(t)
\frac{\partial^3{F}}{\partial{t}
	\partial{z^2}}
+ \kappa_{ttzz}(t)\frac{\partial^4{F}}
{\partial{t^2}\partial{z^2}}
+ \kappa_{tttzz}(t)\frac{\partial^5{F}}
{\partial{t^3}\partial{z^2}}
+\cdots\right)
+\cdots\cdots\Bigg].
\label{convection term of equation of F}
\end{eqnarray}
Inserting Equation
(\ref{convection term of equation of F}) into the convection term
$\kappa_{tz}
\partial^2{F}/\partial{t}\partial{z}$
in Equation (\ref{equation of F})
and neglecting the higher order derivative terms,
we can obtain
another version of
the telegraph equation,
\begin{eqnarray}
\frac{\partial{F}}{\partial{t}}
+\tau'(t)\frac{\partial^2{F}}{\partial{t^2}}
+\kappa_z(t)\frac{\partial{F}}{\partial{z}}
=\kappa_{zz}(t)\frac{\partial^2{F}}
{\partial{z^2}}
\label{new telegraph equation}
\end{eqnarray}
with
\begin{eqnarray}
\tau'(t)=\frac{\kappa_{tz}
(t)}{\kappa_{z}(t)}.
\label{tau}
\end{eqnarray}
We find that the only
difference between Equations (\ref{new telegraph equation})
and (\ref{telegraph equation})
is caused by the fact that
$\tau(t)$ is not equal to $\tau'(t)$.

\section{THE MODIFIED FORMULA OF DIFFUSION COEFFICIENT
WITH DISPLACEMENT VARIANC}
\label{THE PARALLEL DIFFUSION
COEFFICIENT REPRESENTED
BY ALONG-FIELD DISPLACEMENT MOMENTS}
In the following, we derive
the modified formula
of the parallel
diffusion coefficient
with the parallel displacement variance
induced by the adiabatic focusing
(hereafter MDCDV).

\subsection{MDCDV
for the variable coefficient
differential equation}
\label{The parallel diffusion
coefficient of Equation}

By multiplying Equation
(\ref{equation of F})
with
$\langle (\Delta z) \rangle$ and
integrating the result, one can  find
\begin{eqnarray}
\frac{d}{dt}\langle (\Delta z)
\rangle&=&\int_{-\infty}^\infty
dz (\Delta z)\frac{\partial{F}}
{\partial{t}}\nonumber\\
&=&\int_{-\infty}^\infty dz
(\Delta z)\Bigg[\left(-\kappa_z(t)
\frac{\partial{F}}{\partial{z}}
+\kappa_{zz}(t)
\frac{\partial^2{F}}{\partial{z^2}}
+\kappa_{zzz}(t)
\frac{\partial^3{F}}{\partial{z^3}}
+\kappa_{zzzz}(t)
\frac{\partial^4{F}}{\partial{z^4}}
+\cdots+\kappa_{z\cdots z}(t)
\frac{\partial^n{F}}{\partial{z^n}}
+\cdots\right)\nonumber\\
&&+\left( \kappa_{tz}(t)
\frac{\partial^2{F}}
{\partial{t}\partial{z}}
+ \kappa_{ttz}(t)
\frac{\partial^3{F}}
{\partial{t^2}\partial{z}}
+ \kappa_{tttz}(t)
\frac{\partial^4{F}}
{\partial{t^3}\partial{z}}
+\cdots
+\kappa_{t\cdots tz}(t)
\frac{\partial^{n+1}{F}}
{\partial{t^n}\partial{z}}
+\cdots
\right)\nonumber\\
&&+\left( \kappa_{tzz}(t)
\frac{\partial^3{F}}
{\partial{t}\partial{z^2}}
+ \kappa_{ttzz}(t)
\frac{\partial^4{F}}
{\partial{t^2}\partial{z^2}}+
+\cdots+\kappa_{ttzz}(t)
\frac{\partial^4{F}}
{\partial{t^2}\partial{z^2}}
+ \kappa_{t\cdots tzz}(t)
\frac{\partial^{n+2}{F}}
{\partial{t^n}\partial{z^2}}
+\cdots\right)
+\cdots\cdots\Bigg].
\label{1-general}
\end{eqnarray}
To proceed,
we have to obtain the integrals
on the right hand side of
the latter equation
with the following conditions
\begin{eqnarray}
&&F(z=\pm\infty)=0,\\
&&\frac{\partial^n{F}}
{\partial{z^n}}(z=\pm\infty)=0
\hspace{0.2cm} n=1,2,3,\cdots.
\end{eqnarray}
By performing integration
in parts one can obtain
\begin{eqnarray}
&&\int_{-\infty}^\infty
dz (\Delta z)\left(-\kappa_z(t)
\frac{\partial{F}}{\partial{z}}\right)
=-\kappa_z(t)\int_{-\infty}^\infty
dz (\Delta z)\frac{\partial{F}}
{\partial{z}}=\kappa_z(t),
\label{1-z}\\
&&\int_{-\infty}^\infty
dz (\Delta z)\left(\kappa_{zz}(t)
\frac{\partial^2{F}}{\partial{z^2}}\right)
=\kappa_{zz}(t)\int_{-\infty}^\infty
dz (\Delta z)\frac{\partial^2{F}}
{\partial{z^2}}=0.
\label{1-zz}
\end{eqnarray}
We can also find that the integrals
are equal to zero if the order of spatial
derivative of the integrands is
higher than one,
\begin{eqnarray}
\int_{-\infty}^\infty dz (\Delta z)
\left(\kappa_{z\cdots z}(t)
\frac{\partial^n{F}}{\partial{z^n}}\right)
=\kappa_{z\cdots z}(t)\int_{-\infty}^\infty
dz (\Delta z)
\frac{\partial^n{F}}{\partial{z^n}}
=0 \hspace{0.5cm}n=2,3,\cdots.
\label{1-zn}
\end{eqnarray}
It is also easy to find the
following result by integration
by parts
\begin{eqnarray}
\int_{-\infty}^\infty dz
(\Delta z)\left(\kappa_{z\cdots z}(t)
\frac{\partial^{n+m}{F}}{\partial{t^m}
\partial{z^n}}\right)=\kappa_{z\cdots z}(t)
\frac{d^m}{dt^m}\int_{-\infty}^\infty
dz (\Delta z)
\frac{\partial^n{F}}{\partial{z^n}}
=0 \hspace{0.5cm}n=1,2,3,\cdots,
\hspace{0.2cm}m=1,2,3,\cdots.
\label{1-tmzn}
\end{eqnarray}
Inserting Equations (\ref{1-z})
- (\ref{1-tmzn}) into Equation
(\ref{1-general}), one can obtain
the following formula
\begin{equation}
\frac{d}{dt}\langle (\Delta z)
\rangle=\kappa_z(t).
\label{1-result}
\end{equation}

Furthermore, analogous to
Equation (\ref{1-general}),
the formula of second order
moment of the parallel
displacement
can be written as
\begin{eqnarray}
\frac{d}{dt}\langle (\Delta z)^2
\rangle&=&\int_{-\infty}^\infty
dz (\Delta z)^2
\frac{\partial{F}}{\partial{t}}
\nonumber\\
&=&\int_{-\infty}^\infty dz
(\Delta z)^2
\Bigg[\left(-\kappa_z(t)
\frac{\partial{F}}{\partial{z}}
+\kappa_{zz}(t)
\frac{\partial^2{F}}{\partial{z^2}}
+\kappa_{zzz}(t)
\frac{\partial^3{F}}{\partial{z^3}}
+\kappa_{zzzz}(t)
\frac{\partial^4{F}}{\partial{z^4}}
+\cdots+\kappa_{z\cdots z}(t)
\frac{\partial^n{F}}{\partial{z^n}}
+\cdots\right)\nonumber\\
&{}&+\left( \kappa_{tz}
\frac{\partial^2{F}}
{\partial{t}\partial{z}}
+ \kappa_{ttz}(t)
\frac{\partial^3{F}}
{\partial{t^2}\partial{z}}
+ \kappa_{tttz}(t)
\frac{\partial^4{F}}
{\partial{t^3}\partial{z}}
+\cdots
+\kappa_{t\cdots tz}(t)
\frac{\partial^{n+1}{F}}
{\partial{t^n}\partial{z}}
+\cdots\right)\nonumber\\
&&+\left( \kappa_{tzz}(t)
\frac{\partial^3{F}}
{\partial{t}\partial{z^2}}
+ \kappa_{ttzz}(t)\frac{\partial^4{F}}
{\partial{t^2}\partial{z^2}}
+ \kappa_{tttzz}(t)\frac{\partial^5{F}}
{\partial{t^3}\partial{z^2}}
+\cdots+\kappa_{t\cdots tzz}(t)
\frac{\partial^{n+2}{F}}
{\partial{t^n}\partial{z^2}}
+\cdots\right)
+\cdots\cdots\Bigg].
\label{2-general}
\end{eqnarray}
The integrals on the right
hand side of the latter
equation
can be easily found
by part integral
\begin{eqnarray}
&&\int_{-\infty}^\infty dz
(\Delta z)^2\left(-\kappa_z(t)
\frac{\partial{F}}{\partial{z}}
\right)
=-\kappa_z(t)\int_{-\infty}^\infty
dz (\Delta z)^2\frac{\partial{F}}
{\partial{z}}
=2\kappa_z (t)\langle (\Delta z)
\rangle
\label{2-z}\\
&&\int_{-\infty}^\infty dz
(\Delta z)^2\left(\kappa_{zz}(t)
\frac{\partial^2{F}}{\partial{z^2}}
\right)
=\kappa_{zz}(t)\int_{-\infty}^\infty
dz (\Delta z)^2\frac{\partial^2{F}}
{\partial{z^2}}
=2\kappa_{zz}(t)
\label{2-zz}\\
&&\int_{-\infty}^\infty dz
(\Delta z)^2\left(\kappa_{z\cdots z}
(t)\frac{\partial^n{F}}
{\partial{z^n}}\right)
=\kappa_{z\cdots z}(t)
\int_{-\infty}^\infty dz
(\Delta z)^2\frac{\partial^n{F}}
{\partial{z^n}}
=0 \hspace{0.5cm}n=3, 4, 5,
\cdots\\
&&\int_{-\infty}^\infty dz
(\Delta z)^2
\left(\kappa_{t\cdots tz}(t)
\frac{\partial^{n+1}{F}}
{\partial{t^n}\partial{z}}\right)
=\kappa_{t\cdots tz}(t)
\frac{d^n}{dt^n}
\int_{-\infty}^\infty dz
(\Delta z)^2
\frac{\partial{F}}{\partial{z}}
=-2\kappa_{t\cdots tz}(t)
\frac{d^n}{dt^n}\langle
(\Delta z) \rangle
\hspace{0.5cm}n=1,2,3,
\cdots.
\label{1-tzn}\\
&&\int_{-\infty}^\infty dz
(\Delta z)^2
\left(\kappa_{t\cdots tz\cdots z}(t)
\frac{\partial^{n+m}{F}}
{\partial{t^n}\partial{z^m}}\right)
=\kappa_{t\cdots tz\cdots z}(t)
\frac{d^n}{dt^n}
\int_{-\infty}^\infty
dz (\Delta z)^2
\frac{\partial{F^m}}
{\partial{z^m}}=0
\hspace{0.5cm}n=1,2,3,
\cdots, \hspace{0.2cm}m=2,3,
\cdots.
\label{2-tzn}
\end{eqnarray}
With Equations
(\ref{2-z})-(\ref{2-tzn}),
Equation (\ref{2-general})
becomes
\begin{eqnarray}
\frac{d}{dt}\langle (\Delta z)^2
\rangle =2\kappa_{zz}(t)+2\left[\kappa_z(t)
\langle (\Delta z) \rangle
-\kappa_{tz}(t)\frac{d}{dt}
\langle (\Delta z) \rangle
-\kappa_{ttz}(t)\frac{d^2}{dt^2}
\langle (\Delta z) \rangle
-\kappa_{tttz}(t)\frac{d^3}{dt^3}
\langle (\Delta z) \rangle
-\cdots\cdots\right].
\label{2-result}
\end{eqnarray}
The latter equation can be rewritten as
\begin{eqnarray}
\kappa_{zz}(t)=\frac{1}{2}
\frac{d}{dt}\langle
(\Delta z)^2 \rangle
+C(t)
\end{eqnarray}
with
\begin{eqnarray}
C(t)=-\kappa_z(t)\langle (\Delta z)\rangle
+\kappa_{tz}(t)\frac{d}{dt}
\langle (\Delta z) \rangle
+\kappa_{ttz}(t)\frac{d^2}{dt^2}
\langle (\Delta z) \rangle
+\kappa_{tttz}(t)\frac{d^3}{dt^3}
\langle (\Delta z) \rangle
+\cdots\cdots.
\end{eqnarray}
Here, $C(t)$
is related to convection
of energetic particles.

By employing formula $\sigma^2=\langle
(\Delta z)^2 \rangle
-\langle (\Delta z)
\rangle^2$ and Equation
(\ref{1-result})
 we can obtain
\begin{eqnarray}
\kappa_{zz}(t)=\frac{1}{2}
\frac{d\sigma^2}{dt}
+\left[\kappa_{tz}(t)\frac{d}{dt}
\langle (\Delta z) \rangle
+\kappa_{ttz}(t)\frac{d^2}{dt^2}
\langle (\Delta z) \rangle
+\kappa_{tttz}(t)\frac{d^3}{dt^3}
\langle (\Delta z) \rangle
+\cdots\cdots\right].
\label{center formula-1}
\end{eqnarray}
Inserting Equation (\ref{1-result})
into the latter equation one
can find
\begin{eqnarray}
\kappa_{zz}(t)=\frac{1}{2}
\frac{d\sigma^2}{dt}
+\left[\kappa_{tz}(t)
\kappa_z(t)
+\kappa_{ttz}(t)\frac{d}{dt}
\kappa_z(t)
+\kappa_{tttz}(t)\frac{d^2}{dt^2}
\kappa_z(t)+\cdots\cdots\right].
\label{center formula-2}
\end{eqnarray}
The latter equation can
be rewritten as
\begin{eqnarray}
\frac{d\sigma^2}{dt}
=2\kappa_{zz}(t)
-\left[2\kappa_{tz}(t)
\kappa_z(t)
+2\kappa_{ttz}(t)\frac{d}{dt}
\kappa_z(t)+2\kappa_{tttz}(t)
\frac{d^2}{dt^2}\kappa_z(t)
+\cdots\cdots\right].
\label{kzz-0}
\end{eqnarray}

Equation (\ref{center formula-1})
or (\ref{kzz-0}) is the modified DCDV
(MDCDV).
In comparison with the well-known
Equation (\ref{kzz-sigma})
(represented by DCDV),
one can find that MDCDV
is a more general formula.
That is, DCDV
is the special case of MDCDV
when the cross terms with the first
order spatial derivative or
the convection term are
equal to zero.

Because the coefficients
$\kappa_{z}(t)$, $\kappa_{zz}(t)$,
$\kappa_{tz}(t)$, $\cdots\cdots$
are all time dependent,
the relation $\sigma^2\propto t$
does not hold for MDCDV.
Therefore,
the transport
described by MDCDV
is not in diffusion regime.
Since Equation (\ref{kzz-0}) is
derived from Equation
(\ref{equation of F}),
the transport described
by Equation (\ref{equation of F})
is also not in diffusion regime.

\subsection{MDCDV
for the telegraph equation}
\label{The parallel diffusion
coefficient of the telegraph equation}

For the telegraph equation
derived in
\citet{LitvinenkoASchlickeiser2013}
the corresponding first
and second order moments of the
parallel displacement
can be written as, respectively
\begin{eqnarray}
&&\frac{d}{dt}\langle (\Delta z)
\rangle=\kappa_z(t)-\tau(t)
\frac{d^2{}}{d{t^2}}\langle
(\Delta z)  \rangle,
\label{1-result for
telegraph equation}\\
&&\frac{d}{dt}\langle
(\Delta z)^2 \rangle
=2\kappa_z(t) \langle (\Delta z)
\rangle+2\kappa_{zz}(t)
-\tau(t)\frac{d^2}{dt^2}\langle
(\Delta z)^2  \rangle.
\label{2-result for
telegraph equation}
\end{eqnarray}
By combining the latter equations
one can derive with $\sigma^2=\langle
(\Delta z)^2 \rangle -\langle
(\Delta z) \rangle^2$
\begin{eqnarray}
\kappa_{zz}(t)&=&\frac{1}{2}
\frac{d\sigma^2}{dt}
+\tau(t)\Bigg[\frac{1}{2}
\frac{d^2}{dt^2}\langle (\Delta z)^2
\rangle-\langle (\Delta z)
\rangle\frac{d^2}{dt^2}\langle
(\Delta z) \rangle\Bigg].
\end{eqnarray}
Inserting Equation (\ref{tau})
into the latter equation,
we can obtain
\begin{eqnarray}
\kappa_{zz}(t)
=\frac{1}{2}\frac{d\sigma^2}{dt}
-\frac{\kappa_{tzz}(t)}
{\kappa_{zz}(t)}\Bigg[\frac{d^2}{dt^2}
\langle (\Delta z)^2 \rangle-\langle
(\Delta z) \rangle\frac{d^2}{dt^2}
\langle (\Delta z) \rangle\Bigg].
\label{kzz for telegraph equation}
\end{eqnarray}

By subtracting Equation
(\ref{kzz for telegraph equation})
from Equation (\ref{center formula-2})
we can obtain the error formula
\begin{eqnarray}
E(t)
=\left[\kappa_{tz}(t)\kappa_z(t)
+\kappa_{ttz}(t)\frac{d}
{dt}\kappa_z(t)
+\kappa_{tttz}(t)\frac{d^2}{dt^2}
\kappa_z(t)+\cdots\cdots\right]
+\frac{\kappa_{tzz}(t)}
{\kappa_{zz}(t)}\Bigg[\frac{d^2}{dt^2}
\langle (\Delta z)^2 \rangle
-\langle (\Delta z)
\rangle\frac{d^2}{dt^2}\langle
(\Delta z) \rangle\Bigg],
\label{difference}
\end{eqnarray}
which comes from the iteration operation
used in the derivation
of telegraph equation.
In the future, this formula
will be explored in detail.

In addition, Equation
(\ref{1-result for telegraph equation})
can be rewritten as
\begin{eqnarray}
\tau(t)\frac{d^2{}}{d{t^2}}
\langle (\Delta z) \rangle
+\frac{d}{dt}\langle (\Delta z)
\rangle=\kappa_z(t).
\label{11-differential equation}
\end{eqnarray}
If the coefficients $\tau(t)$
and $\kappa_z(t)$ can be determined,
the latter equation might
be solved.
Similarily, Equation
(\ref{2-result for
telegraph equation})
can be rewritten as
\begin{eqnarray}
\tau(t)\frac{d^2}{dt^2}
\langle (\Delta z)^2
\rangle+\frac{d}{dt}\langle
(\Delta z)^2 \rangle =2\kappa_z(t)
\langle (\Delta z) \rangle
+2\kappa_{zz}(t)
\label{22-differential equation}
\end{eqnarray}
Combining the solution of
Equation
(\ref{11-differential equation})
and the coefficients
$\tau(t)$ and $\kappa_z(t)$
with the latter equation,
the formula of
$\langle (\Delta z)^2  \rangle$
might be got.
Some solutions
of Equations (\ref{11-differential equation})
and (\ref{22-differential equation})
will be investigated
in the future.

\subsection{MDCDV for telegraph
Equation (\ref{new telegraph equation})}
\label{The parallel diffusion
coefficient of the new
telegraph equation}

Similarily, for telegraph
Equation (\ref{new telegraph equation}) the first and
second moments can be got
as follows
\begin{eqnarray}
&&\frac{d}{dt}\langle (\Delta z) \rangle
=\kappa_z(t)-\tau'(t)\frac{d^2{}}
{d{t^2}}\langle (\Delta z)  \rangle,\\
&&\frac{d}{dt}\langle (\Delta z)^2
\rangle =2\kappa_z(t) \langle (\Delta z)
\rangle+2\kappa_{zz}(t)
-\tau'(t)\frac{d^2}{dt^2}
\langle (\Delta z)^2  \rangle.
\end{eqnarray}
The corresponding MDCDV becomes
\begin{eqnarray}
\kappa_{zz}(t)
=\frac{1}{2}\frac{d\sigma^2}{dt}
+\frac{\kappa_{tz}(t)}{\kappa_{z}(t)}
\Bigg[\frac{d^2}{dt^2}\langle (\Delta z)^2
\rangle-\langle (\Delta z)
\rangle\frac{d^2}{dt^2}\langle
(\Delta z) \rangle\Bigg].
\end{eqnarray}

\section{SPECIAL CASES}
\label{SPECIAL CASES}

In the following, we investigate
the special cases of MDCDV
(see Equation (\ref{center formula-1})),
which
correspond to the
simplified versions of Equation
(\ref{equation of F}).

\subsection{Case 1:
constant coefficient
differential equation}
\label{For the case D=D(mu)}
If the temporal effect of
the pitch-angle diffusion
coefficient $D_{\mu\mu}$
can be ignored,
i.e.,
$D_{\mu\mu}=D_{\mu\mu}(\mu)$,
all the coefficients in
Equation (\ref{equation of F})
are constant.
Therefore, Equation
(\ref{equation of F})
is simplified as
\begin{eqnarray}
\frac{\partial{F}}{\partial{t}}=
&&\left(-\kappa_z\frac{\partial{F}}
{\partial{z}}+\kappa_{zz}
\frac{\partial^2{F}}{\partial{z^2}}
+\kappa_{zzz}
\frac{\partial^3{F}}{\partial{z^3}}
+\kappa_{zzzz}
\frac{\partial^4{F}}{\partial{z^4}}
+\cdots\right)
+\left( \kappa_{tz}
\frac{\partial^2{F}}{\partial{t}
\partial{z}}
+ \kappa_{ttz}\frac{\partial^3{F}}
{\partial{t^2}\partial{z}}
+ \kappa_{tttz}\frac{\partial^4{F}}
{\partial{t^3}\partial{z}}
+\cdots\right)\nonumber\\
&&+\left( \kappa_{tzz}
\frac{\partial^3{F}}{\partial{t}
\partial{z^2}}
+ \kappa_{ttzz}\frac{\partial^4{F}}
{\partial{t^2}\partial{z^2}}
+ \kappa_{tttzz}\frac{\partial^5{F}}
{\partial{t^3}\partial{z^2}}
+\cdots\right)
+\cdots\cdots.
\label{equation of F with
constant coefficient}
\end{eqnarray}
This is a constant coefficient
linear differential equation
of the isotropic distribution
function.
The first
and second order moments of
the parallel displacement can
be found as
\begin{eqnarray}
&&\frac{d}{dt}\langle (\Delta z)
\rangle=\kappa_z=constant,
\label{1 for constant coefficient}\\
&&\frac{d}{dt}\langle (\Delta z)^2
\rangle =2\kappa_z \langle (\Delta z)
\rangle+2\kappa_{zz}
-2\kappa_{tz}\kappa_z.
\label{2 for constant coefficient}
\end{eqnarray}
By combining the latter two formulas
one can derive
\begin{eqnarray}
\kappa_{zz}=\frac{1}{2}
\frac{d\sigma^2}{dt}
+\kappa_{tz}\kappa_z.
\label{center formula-3}
\end{eqnarray}
We can find
that DCDV
is the special case of
the latter equation
when $\kappa_{tz}\kappa_z=0$.

Because the coefficients
$\kappa_z$, $\kappa_{zz}$
and $\kappa_{tz}$ are all
constants,
Equation (\ref{center formula-3})
can
be rewritten as
\begin{eqnarray}
\frac{d\sigma^2}{dt}=2\kappa_{zz}
-2\kappa_{tz}\kappa_z=constant.
\label{sigma}
\end{eqnarray}
By setting the initial value
of $\sigma^2$ as zero, from
the latter equation we can
obtain
\begin{eqnarray}
\sigma^2=\alpha t.
\label{formula of diffusion regime}
\end{eqnarray}
with
\begin{eqnarray}
\alpha=2\kappa_{tz}\kappa_z
+2\kappa_{zz}=constant
\label{alpha}
\end{eqnarray}
Equation
(\ref{formula of diffusion regime})
describes diffusive regime.
Therefore, for $D_{\mu\mu}
=D_{\mu\mu}(\mu)$
the constant coefficient linear
differential equation of the
isotropic distribution function
(see Equation (\ref{equation of F
with constant coefficient})) can
describe the diffusion
of the energetic particles.
If $\kappa_z$ or $\kappa_{tz}$
can be ignored,
DCDV (Equation (\ref{kzz-sigma}))
is obtained.

\subsection{Case 2:
diffusion convection equation}
\label{For the case diffusion
convection equation}
If Equation
(\ref{equation of F with
constant coefficient})
is simplified to the
diffusion convection equation
\begin{eqnarray}
\frac{\partial{F}}{\partial{t}}
=-\kappa_z\frac{\partial{F}}
{\partial{z}}+\kappa_{zz}
\frac{\partial^2{F}}{\partial{z^2}},
\label{simplified equation of
F with constant coefficient-1}
\end{eqnarray}
Equation (\ref{center formula-3})
becomes DCDV.

\subsection{Case 3:
diffusion equation}
\label{For the case diffusion equation}
If the convection term can be
neglected, the diffusion convection
Equation (\ref{simplified equation of
	F with constant coefficient-1})
can be simplified
to diffusion equation
\begin{eqnarray}
\frac{\partial{F}}{\partial{t}}
=\kappa_{zz}\frac{\partial^2{F}}
{\partial{z^2}}.
\end{eqnarray}
Therefore, Equation
(\ref{1 for constant coefficient})
becomes
\begin{equation}
\frac{d}{dt}\langle (\Delta z) \rangle=0.
\label{1-result for constant ceofficient}
\end{equation}
Consequently,
it is easy to see that
the parallel diffusion coefficient
can be shown as
\begin{eqnarray}
\kappa_{zz}=\frac{1}{2}\frac{d}{dt}
\langle (\Delta z)^2 \rangle .
\label{center formula corresponding
to diffusion equation}
\end{eqnarray}

\subsection{Case 4: the constant
background magnetic field}
\label{For the case: the constant
background magnetic field}
For the constant background magnetic
field, the Fokker-Planck equation with
time-dependent pitch-angle diffusion
coefficient is
\begin{equation}
\frac{\partial{f}}{\partial{t}}
+ v\mu
\frac{\partial{f}}
{\partial{z}}=
\frac{\partial{}}{\partial{\mu}}
\left[D_{\mu \mu}(\mu,t)
\frac{\partial{f}}
{\partial{\mu}}\right].
\label{Fokker-Planck equation
for constant mean magnetic field}
\end{equation}
By using the similar method
in Section 2,
from the latter equation we can
obtain the variable coefficient
differential equation of isotropic
distribution function  as
\begin{eqnarray}
\frac{\partial{F}}{\partial{t}}=
&&\left(\kappa_{zz}(t)
\frac{\partial^2{F}}{\partial{z^2}}
+\kappa_{zzz}(t)
\frac{\partial^3{F}}{\partial{z^3}}
+\kappa_{zzzz}(t)
\frac{\partial^4{F}}{\partial{z^4}}
+\cdots\right)
+\left( \kappa_{tz}(t)
\frac{\partial^2{F}}{\partial{t}\partial{z}}
+ \kappa_{ttz}(t)
\frac{\partial^3{F}}
{\partial{t^2}\partial{z}}
+ \kappa_{tttz}(t)
\frac{\partial^4{F}}{\partial{t^3}
\partial{z}}+\cdots\right)\nonumber\\
&&+\left( \kappa_{tzz}(t)
\frac{\partial^3{F}}{\partial{t}
\partial{z^2}}
+ \kappa_{ttzz}(t)\frac{\partial^4{F}}
{\partial{t^2}\partial{z^2}}
+ \kappa_{tttzz}(t)\frac{\partial^5{F}}
{\partial{t^3}\partial{z^2}}+\cdots\right)
+\cdots\cdots.
\label{equation of F for
constant mean magnetic field}
\end{eqnarray}

Comparing
with Equation (\ref{equation of F}),
we find that Equation
(\ref{equation of F for
constant mean magnetic field})
does not contain the convection
term $\kappa_z(t)
\partial{F}/\partial{z}$
which is caused
by the adiabatic
focusing effect.
The first order moment
of parallel displacement
corresponding to Equation
(\ref{equation of F for
constant mean magnetic field})
is given as,
\begin{eqnarray}
&&\frac{d}{dt}\langle (\Delta z)
\rangle=\kappa_z(t)=0.
\label{1-result for constant
mean magentic field}
\end{eqnarray}
Similarily, we can find the second
order moment of the parallel
displacement as
\begin{eqnarray}
\frac{d}{dt}\langle (\Delta z)^2
\rangle =2\kappa_{zz}(t).
\label{2-result for constant
mean magentic field}
\end{eqnarray}
This equation can be
rewritten as
\begin{eqnarray}
\kappa_{zz}(t)=\frac{1}{2}
\frac{d}{dt}\langle (\Delta z)^2
\rangle.
\label{kzz for constant mean
magentic field}
\end{eqnarray}

\section{SUMMARY AND CONCLUSION}
\label{SUMMARY AND CONCLUSION}

The
parallel diffusion coefficient
can be defined as
one half of the temporal derivative
of variance of the parallel displacement,
DCDV (Equation (\ref{kzz-sigma})).
It is shown that DCDV is only suitable to
the transport equation
without the convection term or
without the cross terms.
If the parallel diffusion coefficient
$\kappa_{zz}$ is a constant,
the formula $\sigma^2=2\kappa_{zz}t$ holds,
which means the transport regime is diffusion.
The purpose of this work is to derive
the formula of the parallel
diffusion coefficient $\kappa_{zz}$
 with the parallel displacement
 variance $\sigma^2$
for the more general transport
equation of the isotropic
distribution function.
We start from the Fokker-Planck equation
with time-dependent pitch-angle
diffusion coefficient $D_{\mu\mu}(\mu,t)$,
which is an extension of the
Fokker-Planck equation with
time-independent
pitch-angle diffusion coefficient.
From this equation,
by employing the
perturbation method which is frequently
used in relevant research
\citep[see, e.g.,][]{BeeckEA1986,
BieberEA1990, SchlickeiserEA2008,
SchlickeiserEA2010,
LitvinenkoASchlickeiser2013,
HeEA2014, WangEA2017b, wq2018},
the variable coefficient differential
equation of the isotropic distribution
 function $F(z,t)$
with the time-dependent coefficients
is derived.
From the equation of $F(z,t)$ the modified
formula describing the relation
of the parallel diffusion coefficient
$\kappa_{zz}$ with the parallel
displacement variance $\sigma^2$
 (MDCDV)
is found.
MDCDV indicates that the parallel
diffusion is determined by not only
the parallel displacement variance
but also all
the coefficients of the cross terms
with the first order spatial derivative
and convection term coefficient.
Because all coefficients are
time-dependent in MDCDV,
the well-known formula $\sigma^2\propto t$
does not hold.
Therefore, the transport of
the energetic charged particles described
by MDCDV
is not in diffusion  regime.

In addition, we study
MDCDV in some special cases.
For the isotropic distribution
function equation with
the constant coefficient,
MDCDV is simplified to
Equation (\ref{center formula-3}).
This formula demonstrates that
the parallel diffusion coefficient
is determined by
the parallel displacement variance,
the convection term coefficient, and
the cross term coefficients with all the first
 order temporal and spatial
derivatives.
Since all the coefficients are constant,
 $\sigma^2\propto t$ can be found.
Therefore, the constant
coefficient differential
equation of  $F(z,t)$ can
describe transport in diffusion
regime.
Furthermore, we explore the
transport equation
with uniform background magnetic field.
For this case, the convection term does
 not occur in the equation of $F(z,t)$.
Accordingly, MDCDV derived
in this paper
reduces to DCDV.
For the diffusion convection
equation and
diffusion equation,
the more simplified equations of
$F(z,t)$,
MDCDV is also reduced to
DCDV.
The telegraph equation is
another frequently used
transport equation,
which is derived by using iteration
method from the isotropic
distribution function equation.
In this article,
we also derive a new telegraph equation,
and explore preliminarily the corresponding
formula of $\kappa_{zz}$
with variance $\sigma^2$.

Instead of adibatic focusing,
solar wind speed is another factor
causing the convection
term in the equation of the isotropic
distribution function.
So, when the phase space equation
with solar
wind speed is projected to real space,
solar wind leads to a convection
term in the real space equation.
We
will explore this topic in the
future.
As a subject of future work we will
perform a
detailed investigation into
the new
telegraph equation.

\acknowledgments

We are partly supported by
grant  NNSFC 41874206, NNSFC 41574172.


\begin{thebibliography}{}
\bibitem[Beeck \& Wibberenz(1986)]{BeeckEA1986}Beeck, J., \& Wibberenz, G. 1986, \apj, 311, 437
\bibitem[Bieber \& Burger(1990)]{BieberEA1990}Bieber, J. W., \& Burger, R. A. 1990, \apj, 348, 597
\bibitem[Earl(1974)]{Earl1974} Earl, J.A. 1974, \apj, 193, 231
\bibitem[Earl(1976)]{Earl1976}Earl, J. A. 1976, \apj, 205, 900
\bibitem[Earl(1981)]{Earl1981}Earl, J. A. 1981, \apj, 251, 739
\bibitem[Fedorov \& Shakhov(2003)]{FedorovShakhov2003}Fedorov, Y. I., \& Shakhov, B. A. 2003, \aap, 402, 805
\bibitem[Effenberger \& Litvinenko(2014)]{EffenbergerEA2014}Effenberger, F., \& Litvinenko, Y. E. 2014, \apj, 783, 15
\bibitem[Hauff \& Jenko(2008)]{HauffEA2008}Hauff, T., \& Jenko, F. 2008, PhPl, 15, 112307
\bibitem[He \& Schlickeiser(2014)]{HeEA2014}He, H.-Q., \& Schlickeiser, R. 2014, \apj, 792, 85
\bibitem[Jokipii(1966)]{Jokipii1966}Jokipii, J. R. 1966, \apj, 146, 480
\bibitem[Kaghashvili et al. (2004)]{KaghashviliEA2004}Kaghashvili, E. Kh., Zank, G. P., Lu, J. Y., \& Dr\"oge, W.   2004, J. Plasma Phys., 7, 505
\bibitem[K\'ota(1994)]{Kota1994}K\'ota, J. 1994, \apj, 427, 1035
\bibitem[K\'ota(2000)]{Kota2000}K\'ota, J. 2000, \jgr, 105, 2403
\bibitem[Kunstmann(1979)]{Kunstmann1979}Kunstmann, J. E. 1979, \apj, 229, 812
\bibitem[Litvinenko(2012a)]{Litvinenko2012a}Litvinenko, Y. E. 2012a, \apj, 752, 16
\bibitem[Litvinenko(2012b)]{Litvinenko2012b}Litvinenko, Y. E. 2012b, \apj, 745, 62
\bibitem[Litvinenko et al. (2015)]{LitvinenkoEA2015}Litvinenko, Y. E., Effenberger, F., \& Schlickeiser, R. 2015, \apj, 806, 217
\bibitem[Litvinenko \& Noble(2016)]{LitvinenkoNoble2016}Litvinenko, Y. E., \& Noble, P. L. 2016, Phys. Plasmas, 23, 062901
\bibitem[Litvinenko \& Schlickeiser(2013)]{LitvinenkoASchlickeiser2013}Litvinenko, Y. E., \& Schlickeiser, R. 2013, \aap, 554, A59
\bibitem[Malkov \& Sagdeev(2015)]{MalkovSagdeev2015}Malkov, M. A., \& Sagdeev, R. Z. 2015, \apj, 808, 157
\bibitem[Matthaeus et al.(1995)]{MatthaeusEA1995}Matthaeus, W. H., Gray, P. C., Pontius, D. H., Jr, \& Bieber, J. W. 1995, \prl, 75, 2136
\bibitem[Matthaeus et al.(2003)]{MatthaeusEA2003}Matthaeus, W. H., Qin, G., Bieber, J. W., \& Zank, G. P. 2003, \apj, 590, L53
\bibitem[Pauls \& Burger(1994)]{PaulsBurger1994}Pauls, H. L., \& Burger, R. A. 1994, \apj, 427, 927
\bibitem[Porr\`a \& Masoliver(1997)]{PorraMasoliver1997}Porr\`a, J. M., \& Masoliver, J. 1997, \pre, 55, 7771
\bibitem[Qin(2007)]{Qin2007}Qin, G. 2007, \apj, 656, 217
\bibitem[Qin et al. (2002a)]{QinEA2002a}Qin, G., Matthaeus, W. H., \& Bieber, J. W. 2002a, \grl, 29, 1048
\bibitem[Qin et al. (2002b)]{QinEA2002b}Qin, G., Matthaeus, W. H., \& Bieber, J. W. 2002b, \apjl, 578, L117
\bibitem[Qin \& Zhang(2014)]{QinEA2014}Qin, G., \& Zhang, L.-H. 2014, \apj, 787, 12
\bibitem[Reimer \& Shalchi (2016)]{ReimerEA2016}Reimer, A., \& Shalchi, A. 2016, \mnras, 456, 3803
\bibitem[Roelof(1969)]{Roelof1969}Roelof, E. C. 1969, in Lectures in High Energy Astrophysics, ed. H. \"Ogelmann \& J. R. Wayland(NASA SP-199: Washington, DC: NASA), 111
\bibitem[Schlickeiser(2002)]{Schlickeiser2002}Schlickeiser, R. 2002, Cosmic Ray Astrophysics (Berlin: Springer)
\bibitem[Schlickeiser et al(2007)]{SchlickeiserEA2007}Schlickeiser, R., Dohle, U., Tautz, R.C., \& Shalchi, A. 2007, \apj, 661, 185
\bibitem[Schlickeiser \& Jenko(2010)]{SchlickeiserEA2010}Schlickeiser, R., \& Jenko, F. 2010, J. Plasma Phys., 76, 317
\bibitem[Schlickeiser \& Shalchi(2008)]{SchlickeiserEA2008}Schlickeiser, R., \& Shalchi, A. 2008, \apj, 686, 292
\bibitem[Schwadron \& Gombosi(2010)]{SchwadronEA2010}Schwadron, N. A., \& Gombosi, T. I. 1994, \jgr, 99, 301
\bibitem[Shalchi(2005)]{Shalchi2005}Shalchi, A. 2005, Phys. Plasmas, 12, 052905
\bibitem[Shalchi(2009a)]{Shalchi2009a}Shalchi, A. 2009a, Nonlinear Cosmic Ray Diffusion Theories, Astrophysics and Space Science Library, Vol. 362 (Berlin: Springer)
\bibitem[Shalchi(2009b)]{Shalchi2009b}Shalchi, A. 2009b, J. Phys. G: Nucl. Part. Phys., 36, 025202
\bibitem[Shalchi(2010)]{Shalchi2010}Shalchi, A. 2010, ApJL, 720, L127
\bibitem[Shalchi(2011)]{Shalchi2011}Shalchi, A. 2011, \apj, 728, 113
\bibitem[Shalchi et al.(2006)]{ShalchiEA2006}Shalchi, A., Bieber, J. W., Matthaeus, W. H., \& Schlickeiser, R. 2006, \apj, 642, 230
\bibitem[Shalchi \& Danos(2013)]{ShalchiEA2013}Shalchi, A., \& Danos, R. J. 2013, \apj, 765, 153
\bibitem[Shalchi \& Kourakis(2007)]{ShalchiAKourakis2007}Shalchi, A., \& Kourakis, I. 2007, Phys. Plasmas, 14, 092903
\bibitem[Shalchi \& Qin(2010)]{ShalchiAQin2010} Shalchi, A., \& Qin, G. 2010, Ap\&SS, 330, 279
\bibitem[Shalchi \& Schlickeiser(2005)]{ShalchiEA2005}Shalchi, A., \& Schlickeiser, R. 2005, \apj, 626, L97
\bibitem[Shalchi et al. (2009a)]{ShalchiEA2009a}Shalchi, A., \v Skoda, T., Tautz, R. C., \& Schlickeiser, R. 2009a, \aap, 507, 589
\bibitem[Shalchi et al. (2009b)]{ShalchiEA2009b}Shalchi, A., \v Skoda, T., Tautz, R. C., \& Schlickeiser, R. 2009b, \prd, 80, 023012
\bibitem[Wang \& Qin(2018)]{wq2018}Wang, J.-F., Qin, G. 2018, \apj, 868, 139
\bibitem[Wang et al.(2017a)]{WangEA2017a}Wang, J.-F., Qin, G., Ma, Q.-M., Song, T., \& Yuan, S.-B. 2017a,  Phys. Plasmas, 24, 082901
\bibitem[Wang et al.(2017b)]{WangEA2017b}Wang, J.-F., Qin, G., Ma, Q.-M., Song, T., \& Yuan, S.-B. 2017b, \apj, 845, 112
\bibitem[Wang \& Qin(2016)]{WangEA2016}Wang, Y., \& Qin, G. 2016, \apj, 820, 61
\bibitem[Zank et al. (2000)]{ZankEA2000}Zank, G. P., Lu, J. Y., Rice, W. K. M., \& Webb, G. M. 2000, J. Plasma Phys., 64, 507

\end{thebibliography}
\end{document}